# Influence of interface dipole layers on the performance of graphene field effect transistors

*Naoka Nagamura[1,2], Hirokazu Fukidome[3,*], Kosuke Nagashio[4], Koji Horiba[5], Takayuki Ide[3], Kazutoshi Funakubo[3], Keiichiro Tashima[3], Akira Toriumi[4], Maki Suemitsu[3], Karsten Horn[6], and Masaharu Oshima[7]*

[1]National Institute for Materials Science, 1-2-1 Sengen, Tsukuba, Ibaraki 305-0047, Japan

[2]PRESTO, Japan Science and Technology Agency, 4-1-8 Honcho, Kawaguchi, Saitama 332-0012, Japan

[3]Research Institute of Electrical Communication, Tohoku University, 2-1-1 Katahira, Aoba-ku, Sendai 980-8577, Japan

[4]Department of Materials Engineering, Graduate School of Engineering, The University of Tokyo, 7-3-1 Hongo, Bunkyo-ku, Tokyo 113-8656, Japan

[5]Photon Factory, Institute of Materials Structure Science, High Energy Accelerator Research Organization, 1-1 Oho, Tsukuba, Ibaraki 305-0801, Japan

[6]Department of Physical Chemistry, Fritz Haber Institute of the Max Plank Society, 14195 Berlin, Germany

[7]Synchrotron Radiation Research Organization, The University of Tokyo, 7-3-1 Hongo, Bunkyo-ku, Tokyo 113-8656, Japan

*Corresponding authors
Email: fukidome@riec.tohoku.ac.jp; Tel: +81-22-217-5484 (Hirokazu Fukidome)

**Abstract**

The linear band dispersion of graphene's bands near the Fermi level gives rise to its unique electronic properties, such as a giant carrier mobility, and this has triggered extensive research in applications, such as graphene field-effect transistors (GFETs). However, GFETs generally exhibit a device performance much inferior compared to the expected one. This has been attributed to a strong dependence of the electronic properties of graphene on the surrounding interfaces. Here we study the interface between a graphene channel and $SiO_2$, and by means of photoelectron spectromicroscopy achieve a detailed determination of the course of band alignment at the interface. Our results show that the electronic properties of graphene are modulated by a hydrophilic $SiO_2$ surface, but not by a hydrophobic one. By combining photoelectron spectromicroscopy with GFET transport property characterization, we demonstrate that the presence of electrical dipoles in the interface, which reflects the $SiO_2$ surface electrochemistry, determines the GFET device performance. A hysteresis in the resistance vs. gate voltage as a function of polarity is ascribed to a reversal of the dipole layer by the gate voltage. These data pave the way for GFET device optimization.

**1. Introduction**

Interfaces of a graphene channel, such as those with gate oxides or contact metals, demand precise and accurate control of electronic level alignment. In graphene, the linear band dispersion near the Fermi level in principle provides excellent intrinsic electronic properties, e.g. an extremely high mobility of carriers, derived from their zero-effective mass. These intrinsic properties make graphene a promising material for next-generation electronics applications, such as graphene-based field-effect transistors (GFETs) and high-electron mobility transistors (HEMTs). At present, however, actual graphene channels used in GFETs actually exhibit transport properties inferior to those anticipated from the intrinsic electronic properties[1-4]. This inferiority is ascribed to the high susceptibility of the electronic properties of graphene to the surrounding interfaces as well as to technological immaturity in graphene device fabrication. This high susceptibility is also in part due to the linear band dispersion. In conventional semiconductors, the average kinetic energy per electron is $E_K \sim \hbar^2 n_d^{2/d}/2m^*$ where $m^*$ is the effective mass and $n_d$ is the average electron density in $d$ spatial dimensions. The Coulomb energy per electron is of the order $E_C \sim e^2 n_d^{1/d}/\varepsilon_0$ where $e$ is the electron charge. Therefore, the effective coupling constant $\alpha_{eff}$, which indicates the ratio of Coulomb over kinetic energy and is related to the strength of the electron-electron interactions, is given by $\alpha_{eff} = E_C/E_K = 2m^* e^2 n_d^{-1/d}/\hbar^2 \varepsilon_0$. This depends on carrier density. On the other hand, the average kinetic energy per electron in graphene is of the order $E_K \sim \hbar v_F n^{1/2}$ where $v_F$ is the Fermi-Dirac velocity and $n$ is the 2D electronic density, owing to its linear dispersion. So, the $\alpha_{eff}$ in 2D graphene is described as $\alpha_{eff} = E_C/E_K = (e^2/\varepsilon_0)/\hbar v_F$. This is independent of the electronic density, but affected by the dielectric constant of the surrounding environment[5-7].

Of the various interfaces, the interface with gate oxides is of particular concern[8,9] because oxide films are the most popular materials for insulating layers in semiconductor devices. The interface between a graphene channel and a gate oxide not only acts as scattering centers for carriers but also causes drastic changes in electronic characteristics of graphene such as e-e interactions described above. Changes in the dielectric constants of gate oxides, such as $SiO_2$, thus influence the transport properties or in other words, the device performance[10]; for example, depositing ice on a graphene channel enhances carrier mobility in the channel[11]. This is explained by a reduction in $\alpha_{eff}$ by the high dielectric constant of ice. Hence the interface chemistry also influences the transport properties. This is shown by reports that a graphene channel interfaced with hydrophilic $SiO_2$ exhibits degraded transport properties, such as a reduced carrier mobility and hysteresis in the resistance-gate bias curve in the gate-bias sweep direction, when compared to a channel interfaced with hydrophobic $SiO_2$[12,13]. The difference between hydrophobic and hydrophilic $SiO_2$ lies in the presence of adsorbed water molecules on the latter, which are sandwiched between graphene and hydrophilic $SiO_2$[14].

The impact of interface physics and chemistry on graphene channels should thus be fully understood for further development in the GFET technology. Imaging techniques are most useful to extract interface characteristics and microstructures[15-18]. To this end, we have developed a core-level photoelectron spectromicroscopy technology, called "3D nano-ESCA," where we can scan the sample with a high lateral spatial resolution (70 nm)[19] to record photoelectron spectra (the "third dimension") to quantitatively analyze electronic level information, such as the Fermi level in graphene, and also chemical states, for example the oxidation valency of $SiO_2$ at the desired points, from core level line positions[20]. We have demonstrated in our previous reports that 3D nano-ESCA is useful for the microscopic investigation of GFETs[21,22]. In fact, microscopic spatial variations of the potential landscape in a GFET were elucidated by measuring the Fermi level using 3D nano-ESCA. Thus, 3D nano-ESCA is a most suitable tool to analyze electronic states, chemical states and, indirectly, transport properties.

This study describes how $SiO_2$ surface chemistry, i.e., hydrophilicity vs. hydrophobicity, modulates the electronic states of the graphene channel from a microscopic viewpoint and then compares the influence of states with the transport properties obtained from the macroscopic electrical characteristics of GFETs.

## 2. Methods

### 2.1. Sample Preparation

Exfoliated graphene was transferred onto $SiO_2$ thin films (90 nm) on $p^+$-Si(100) substrates. The color contrasts in optical images depending on the layer number were emphasized at the graphene sheets on 90 nm $SiO_2$/Si substrates, so the presence of a mono-layer of graphene was confirmed by the optical contrasts and Raman spectroscopy[23]. To prepare a hydrophobic $SiO_2$ thin film, we performed the so-called reoxidation process of the $SiO_2$ thin film by annealing it at 1273 K for 5 min in a 100% oxygen gas flow[13]. This process induces the desorption of $H_2O$ molecules from the surface and produces surface siloxane groups. On the other hand, to prepare hydrophilic $SiO_2$ thin films, an $O_2$ plasma treatment with an $O_2$/Ar mixture (1:9) flow rate of 50 $cm^3$/min was carried out[13]. After the exfoliation of monolayer graphene on the prepared $SiO_2$ thin films, Ni contact electrodes were prepared by vacuum evaporation, and structured by electron-beam lithography. The post annealing procedure was not adopted, and the measurements for sample characterization were performed on as-fabricated devices.

## 2.2. Sample Characterization

Spatially resolved C $1s$, Si $2p$, and O $1s$ core-level photoelectron spectra measurement was carried out using the 3D nano-ESCA instrument installed at the University of Tokyo outstation beamline, BL07LSU at SPring-8[19,24] In this beamline, the synchrotron radiation (SR) beam has a high energy-resolving power ($E/\Delta E > 10^4$). The photon energy of the SR beam used for measurements was 1000 eV. The lateral spatial resolution, i.e., the spot size of the X-rays focused using a Fresnel zone plate, was 70 nm. The energy resolution of the spectrometer was set to 300 meV and the accuracy of the angle resolution was 0.9°. The binding energy scale was calibrated using the photoelectron peaks of a gold mesh foil (Au $4f\,7/2$, binding energy: 84.0 eV) at the same potential as the source electrode, and the Fermi levels detected in valence spectra on Ni electrodes. Details of the experimental setup can be found elsewhere[19,22]. The resistance-gate voltage characteristics were evaluated in ambient air conditions using a semiconductor parameter analyzer (B1500A, Keysight Technologies Inc.).

## 3. Results and Discussion

To quantitatively analyze the impact of $SiO_2$ surface chemistry on band level alignment, we first demonstrate the applicability of 3D nano-ESCA; we then discuss the influence of interface chemistry between graphene and $SiO_2$ on the channel performance. In section 3.3, we compare the electronic states with the device performance (e.g., hysteresis), which varies with $SiO_2$ surface chemistry, and finally in section 3.4, we show that $SiO_2$ surface chemistry affects the electronic states of graphene near metal contacts as well.

### 3.1. 3D nano-ESCA Imaging of GFET

3D nano-ESCA, as illustrated in Fig. 1(a), is used to analyze the electronic structure of a GFET and to quantitatively clarify the effect of surface chemistry of $SiO_2$ thin films on the graphene channel. A GFET structure on a 1 cm × 1 cm substrate is mounted on a sample holder as shown in Fig. 1(b). Electrodes, including source, drain, and gate, are connected to the chamber ground. The optical micrograph of the GFET device structure consisting of a graphene flake channel region and contact metal electrodes is shown in the upper picture in Fig. 1(c). The faint shape of the graphene flake is barely visible. On the other hand, highly spatially resolved elemental mapping of the GFET device, where the intensities of the C 1$s$, Si 2$p$, and Ni 3$p$ core-level spectra are red, green, and pink, respectively, are shown in the lower picture of Fig. 1(c), which is the same region as the upper one. A sharp image is obtained by using the nano-focused X-ray beam (70 nm) with a Fresnel zone plate[19]. 3D nano-ESCA thus has a high enough lateral spatial resolution to reflect the GFET architecture. As discussed above, hydrophobic $SiO_2$ thin films deposited on Si(100) substrates were subjected to the so-called reoxidation process, which leads to the surface being covered by siloxane groups (Fig. 1(d)), while the hydrophilic one is covered with silanol groups (Fig. 1(e))[13,22]. The cross sections of both devices are schematically shown in Figs. 1(d) and 1(e), respectively.

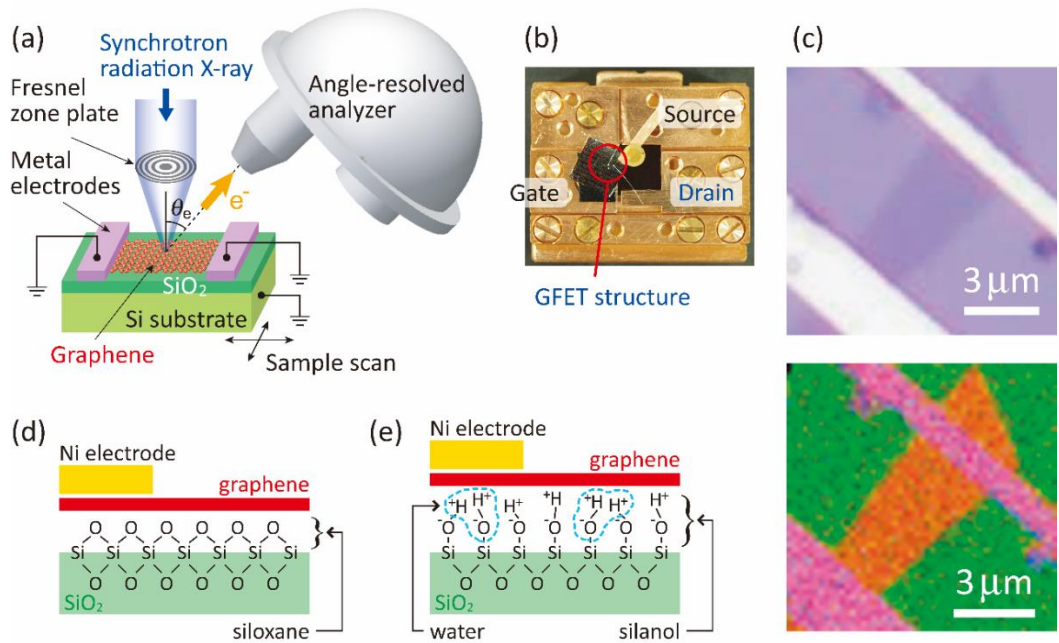

**Figure 1 | Schematic of the measurements and graphene-oxide interface in GFET. (a)** Schematics of the 3D nano-ESCA imaging system. (**b**) Photo of a sample holder upon which a GFET device structure on a substrate is mounted. (**c**) (upper) Optical micrograph of a GFET, barely showing the graphene flake. (lower) Elemental mapping of a GFET using 3D nano-ESCA, where green is the silicon substrate, pink are the nickel contacts, and red the graphene flake. (**d**, **e**) Schematic cross-sections of Ni electrodes and graphene on (d) hydrophobic and (e) hydrophilic SiO₂ thin films, respectively, indicative of the chemical composition.

### 3.2. Interface Chemistry of Graphene with SiO₂

The electronic and chemical states at the interfaces of graphene channels in GFETs with hydrophobic or hydrophilic SiO₂ films were examined by performing a point-for-point spectroscopic analysis of the C 1*s*, Si 2*p*, and O 1*s* core levels at the center of the graphene channels with 3D nano-ESCA, as shown in Figs. 2 and 3.

Along the graphene channels, we used the core-level binding energies to investigate the potential level alignment of graphene on the hydrophobic (blue curves in Figs. 2 and 3) and hydrophilic (red curves in Figs. 2 and 3) $SiO_2$ thin films on the Si(100) substrates. The C 1$s$ spectrum of graphene on the hydrophobic $SiO_2$ thin film on Si(100) has a higher binding energy, compared to that on the hydrophilic $SiO_2$ thin film on Si(100) (Fig. 2(a)). The C 1$s$ spectra can be decomposed into two components by precisely examining the binding energy, which directly reflects the Fermi level position relative to the Dirac point, of the graphene channels (Fig. 2(b)). The lower binding energy component is attributed to graphene, while the higher binding energy component, which is somewhat broader, is attributed to contaminations probably arising from lithographic processing, according to our previous angle-resolved analysis of the C 1$s$ spectra of GFETs[22]. Although these contaminants could have an influence as p-type dopants and scattering centers in the graphene channels[25], the amount of residual carbon contaminants is almost the same between graphene channels on hydrophilic $SiO_2$ and hydrophobic $SiO_2$ according to the intensity of peak components assigned to contaminants in Fig. 2(b), so we assume that the effect of contaminants is the same on the hydrophilic and hydrophobic $SiO_2$. It is obvious that the peak of graphene on a hydrophilic $SiO_2$ thin film has a lower binding energy than that on a hydrophobic $SiO_2$ thin film. This result can be explained by the fact that doping induces a shift in the Fermi level, resulting in a shift in the C 1$s$ binding energy[26], as schematically shown in Fig. 2(c). The binding energy of graphene on a hydrophobic $SiO_2$ thin film is 284.45 eV, which is very close to that of neutral graphene[27,28]. Graphene on a hydrophobic $SiO_2$ thin film on Si(100) therefore exhibits negligible doping; on the other hand, it can be inferred from the lower binding energy (284.23 eV) of the C 1*s* peak that graphene on a hydrophilic SiO$_2$ thin film is hole-doped. The difference between the spectra is about 0.22 eV. Assuming a linear band dispersion of graphene with respect to the wave vector, we use the following equation to estimate the concentration of doped holes ($N_h$) as follows[29,30],

$$E_{DP} - E_F = \hbar v_F \sqrt{\pi} \sqrt{N_h} \qquad (1)$$

Here, $E_F$, $E_{DP}$, $\hbar$, and $v_F$ are the Fermi level, Dirac point energy, reduced Planck's constant, and the Fermi velocity of electrons in graphene (~1.1 × 10$^6$ m/s), respectively. Inserting our experimental value of 0.22 eV as the value of ($E_{DP}$–$E_F$), the concentration of the doped holes ($N_h$) in graphene on hydrophilic SiO$_2$ thin films is estimated to be 2.4 × 10$^{12}$ cm$^{-2}$. The SiO$_2$ surface chemistry thus has a strong influence on the Fermi level position relative to the Dirac point, i.e., the doping strength, in the graphene channel.

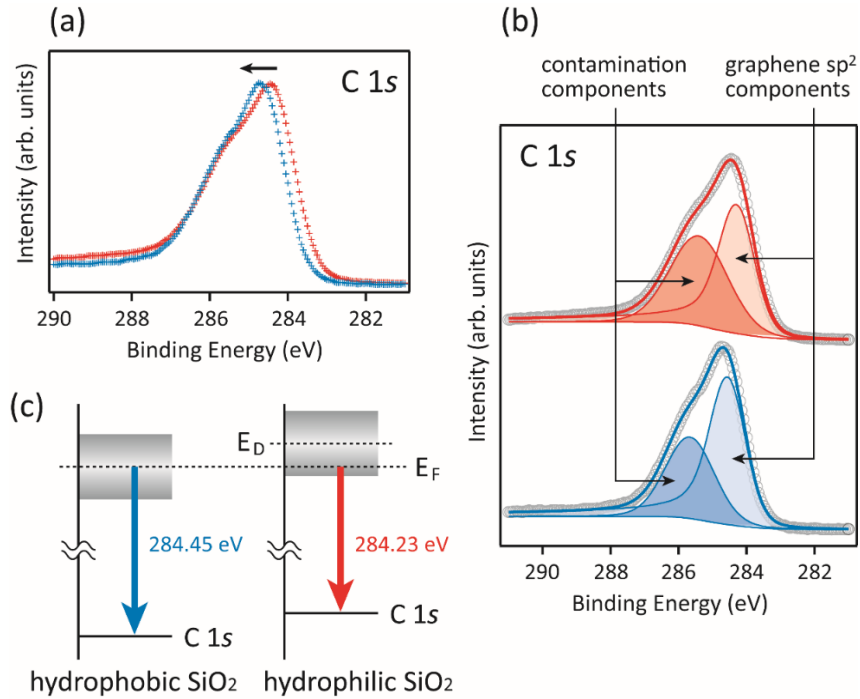

**Figure 2 | Electronic states of graphene channels in contact with SiO$_2$ gate**

**oxides.** (**a**) Pinpoint C 1*s* core-level spectra recorded at the center of the graphene channels on hydrophobic (blue) and hydrophilic (red) $SiO_2$ thin films on Si(100) substrates. (**b**) Decomposition of the spectra displayed in (**a**). (**c**) Schematic diagram to explain the peak shift in graphene at different $SiO_2$ surface conditions. For comparison, the data on hydrophilic $SiO_2$ thin films was sourced from[22].

In order to investigate the influence of $SiO_2$ surface chemistry on the electronic states of graphene, pinpoint measurements of the Si 2*p* and O 1*s* core lines of the $SiO_2$ thin films underneath graphene were conducted at the center of the graphene channels, as shown in Figs. 3(a) and (b), respectively, with the decomposition of these core lines shown in Figs. 3(c) and (d). The Si 2*p* core lines of hydrophobic $SiO_2$ contain a small shoulder around 103 eV, which is not observed in hydrophilic $SiO_2$. This shoulder, which occurs at lower binding energies, is ascribed to the surface siloxane, as schematically shown in Fig. 1, which has a lower valency than stoichiometric $SiO_2$[31]. This is corroborated by the angle dependence of the peak intensity ratio of the siloxane peak over the bulk $SiO_2$ peak (Figs. 4(a) and (b)); the results indicate that siloxane is present on the surface. Both the O 1*s* and Si 2*p* core lines of the hydrophilic $SiO_2$ thin film are shifted towards lower binding energies by a considerable amount (~1.2 eV) when compared to the shifts in their counterparts corresponding to the hydrophobic $SiO_2$ thin film. This means that there is an interfacial layer that affects the binding energies.

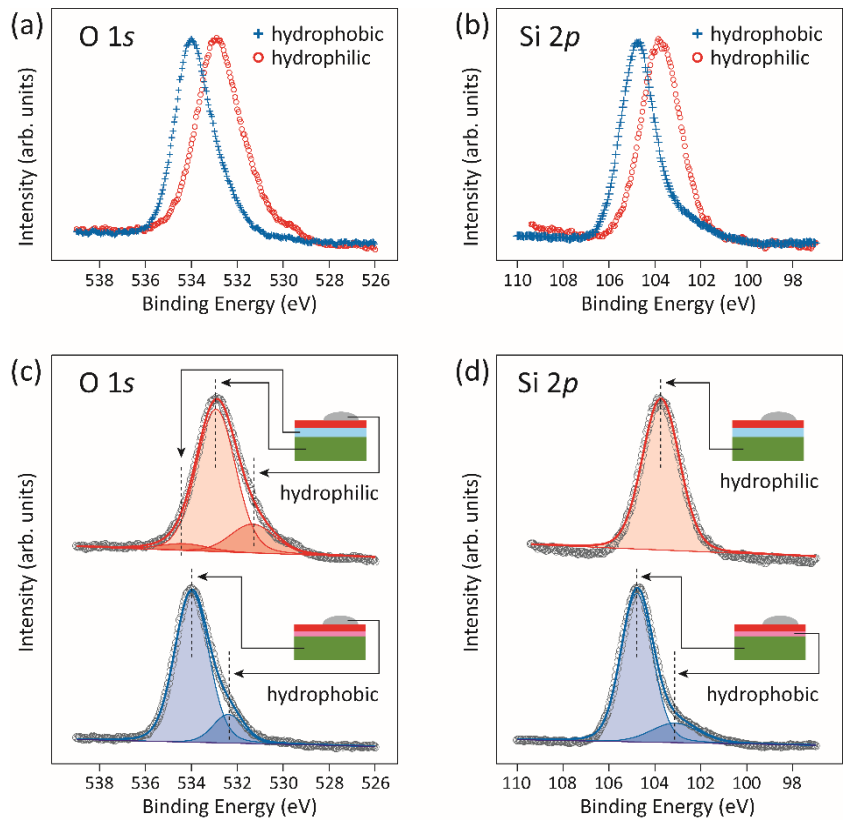

**Figure 3 | Electronic and chemical states of the SiO₂-graphene channel interface.** (**a, b**) Si 2*p* and O 1*s* core level spectra at the center of the graphene channels on hydrophobic (blue plus sign) and hydrophilic (red circle) SiO₂ thin films on Si(100) substrates, respectively. For comparison, the data of a hydrophilic SiO₂ thin films is sourced from ref. 22. (**c, d**) Decomposed Si 2*p* and O 1*s* spectra, respectively.

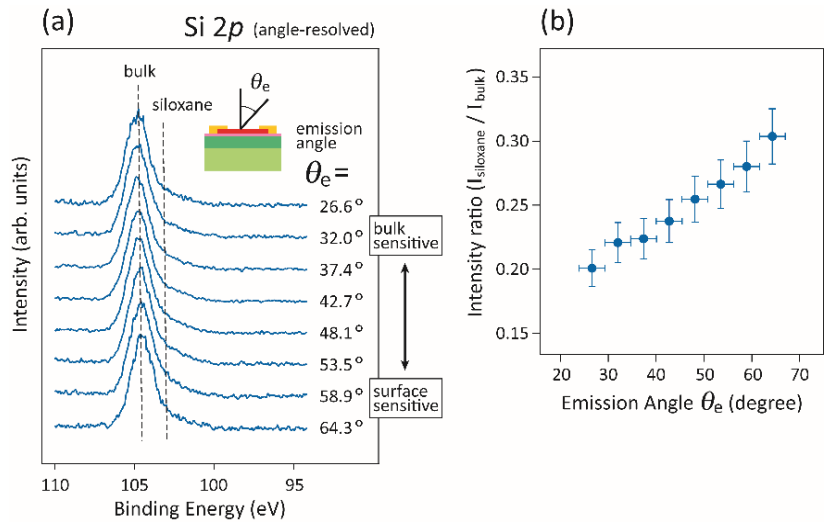

**Figure 4 | Emission angle dependence of surface siloxane and bulk SiO₂.** (**a**) Si 2*p* core level spectra measured in various emission angles. Photoelectron signal from grazing angle is surface sensitive, and that from near normal angle contains information of buried bulk regions. (**b**) Emission angle dependence of the intensity ratio between the shoulder peak at 103 eV derived from the surface siloxane and the main peak component at 105 eV derived from the bulk SiO₂ substrate.

We interpret the shift towards lower binding energies as being not of chemical origin, they would be too large anyway, but due to the potential alignment in the GFET device. The difference in binding energies is very similar ( the width of the O 1*s* peak is slightly different and there is a chemical shift which reveals itself by the presence of a second component in both interfaces). The considerable shift in binding energies can be analyzed using the schematic band diagram of graphene/SiO₂/Si interfaces in GFETs (Fig. 5(a)). In our pinpoint analysis, the graphene channel, metal contacts, and back gate (Si substrate) are grounded, i.e., the applied gate bias ($V_G$) = 0 V, as shown in Fig. 1(b). The Fermi level then extends through all three materials as a straight line. Band alignment at the Si/SiO$_2$ interface can be derived from the Fermi level at the Si surface[20]. SiO$_2$, which has a large bandgap ($\sim$ 9 eV), causes a large potential drop in the Si substrate at the interface, as shown in Fig. 5(a).

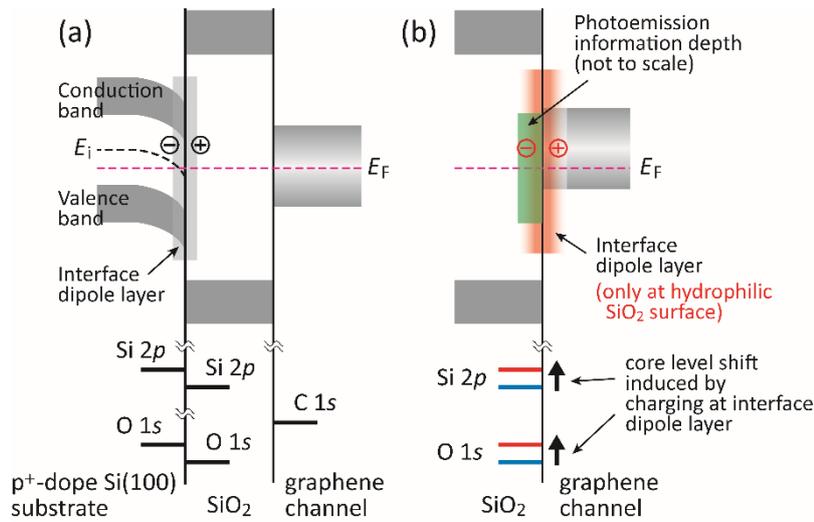

**Figure 5 | Schematic band diagrams of interfaces in GFET structures.** (**a**) Schematic band diagram across the GFET. (**b**) Schematic band diagram of graphene and SiO$_2$ thin films to demonstrate the influence of SiO$_2$ surface chemistry.

A comparison of graphene/hydrophobic SiO$_2$ and graphene/hydrophilic SiO$_2$ interfaces, in terms of the binding energies of the core levels and valence band, is shown in Fig. 5(b). Here, the Fermi level of SiO$_2$ is again aligned with that of graphene. We could detect the graphene, the topmost layer of the SiO$_2$ thin film under graphene, and the graphene/SiO$_2$ interface because the probing depth was a few nanometers, considering the escape depth of photoelectrons with a kinetic energy of about 500 (O 1$s$) and 900 (Si 2$p$) eV[32] at the incident photon energy 1000 eV. The binding energies of the Si $2p$ and O $1s$ core levels thus reflect the changes in valence band alignment with respect to the Fermi level. The point to be noted is the presence or absence of a dipole layer at the graphene/SiO$_2$ interface, which depends on the SiO$_2$ surface chemistry. With respect to the graphene/hydrophilic SiO$_2$ interface, graphene and SiO$_2$ are charged positively and negatively, respectively, as can be inferred from the shifts in the C $1s$ (Fig. 2(a)), Si $2p$ (Fig. 3(a)), and O $1s$ core levels (Fig. 3(b)). We attribute the existence of silanol groups on the surface of the hydrophilic SiO$_2$ thin film to the negative charges on the surface. According to previous theoretical predictions, neither silanol nor siloxane groups cause doping in graphene[33]. While this prediction awaits experimental confirmation, the potential shift due to silanol groups, when in contact with water molecules, may be causing the doping in graphene. In fact, this suggestion is supported by the low value of the acid-dissociation constant (pK$_a$) of the SiO$_2$ surface ($\sim$ 4.5)[34], which indicates a negative charge by the process of giving up a proton in water, which has a higher value pK$_a$ (pH) of 7. This has been verified by in-situ electrochemical Fourier transform infrared spectroscopy (FTIR) in combination with quantum chemical calculations, which indicate that negatively charged silanol groups are formed when a SiO$_2$ surface is in contact with water[35]. Furthermore, our suggestion is supported by previous theoretical studies pointing out the role of water in the doping of graphene on substrates such as SiO$_2$[36,37]. The dipole layer thus consists of positively-charged graphene and a negatively-charged hydrophilic SiO$_2$ thin film, resulting in a potential drop in the layer, as shown in Fig. 5(b). The potential drop shifts the energy position of the Si $2p$ and O $1s$ core lines upwards relative to the Fermi level. In the case of the hydrophobic SiO$_2$ thin film, no dipole layer is present at the interface, hence graphene is not doped, (see Fig. 2), when the surface of the hydrophobic $SiO_2$ thin film is covered with uncharged siloxane groups. A negligible dipole layer is then formed at the interface. This results in a negligible potential drop, as expressed by the straight line across the interface (Fig. 5(b)). Thus, pinpoint core level spectroscopy demonstrates that the $SiO_2$ surface chemistry has a great impact on the interfacial electronic level alignment between graphene and $SiO_2$.

### 3.3. Influence of SiO₂ Surface Chemistry on GFET Electrical Characteristics

The above difference in level alignment in graphene on a hydrophilic or hydrophobic $SiO_2$ substrate is expected to have a strong influence on GFET electrical characteristics as well[12,13], because the electronic states of the graphene channel determine the GFET electrical characteristics. Therefore, we compared the resistance ($R$)-gate voltage ($V_G$) curves of GFETs using hydrophobic and hydrophilic $SiO_2$ thin films as the gate oxides, as shown in Fig. 6(a). In the GFETs, Ni thin films and $p^+$-Si(100) substrates are used as the source/drain electrodes and back gate, respectively. The most striking feature in these curves is the large hysteresis found in the $R$-$V_G$ curve of the GFET using a hydrophilic $SiO_2$ thin film as the gate oxide, but not in the GFET using hydrophobic $SiO_2$[12,13]. The curves for forward and backward sweep on hydrophobic $SiO_2$ are identical and are thus not resolved in Fig. 6(a). Such hysteresis, which is reproduced over many consecutive sweeps[13], indicates that the doping type changes with a change in the direction of the gate voltage sweep. Because the dipole layer induces a difference in the level alignment between the two GFETs (Fig. 5(b)), it is obvious that the dipole layer formed between graphene and the hydrophilic $SiO_2$ thin film affects the doping level of the graphene channel, as schematically shown in Fig. 6(b). What we observe here is that the reversal of gate voltage inverts the polarity of the dipole layer, which arises from the polarity inversion of the charging states of graphene and the hydrophilic $SiO_2$ surface. As an aside we note that the consumption of gate voltage by the dipole layer (Fig. 6(b)) can cause a broadening in the width of the $R$-$V_G$ curve of the GFET with hydrophilic $SiO_2$ rather than hydrophobic $SiO_2$[38]. This is because the effective gate capacitance $C$ increases by the dipole layer in the formula[39] which represents the graphene resistance $R$;

$$R = \frac{L}{w} \times \frac{1}{e\mu\sqrt{n_0^2 + (C|V_G - V_D|)^2}} + R_C \, , \qquad (2)$$

where $L/w$ is the aspect ratio of the transistor, $\mu$ is the mobility, $n_0$ is the residual charge, and $R_C$ is the constant background resistance.

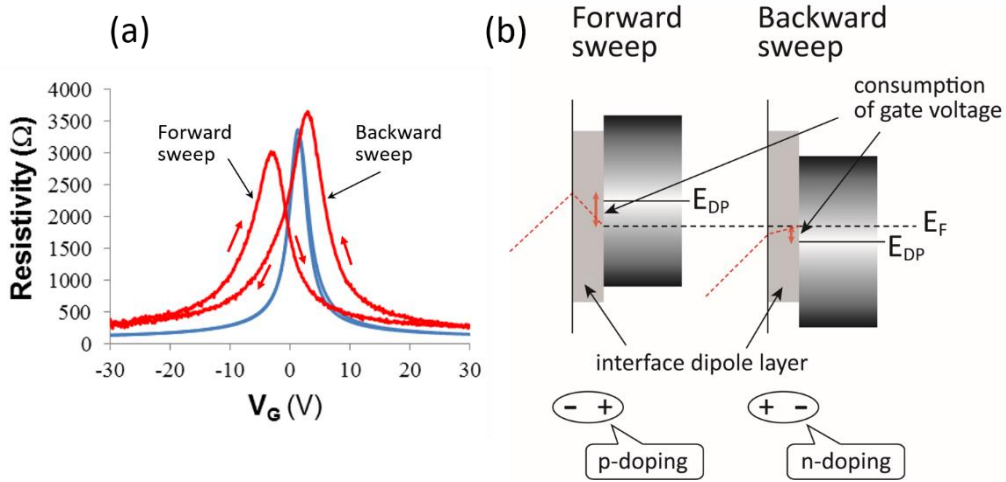

**Figure 6 | Influence of $SiO_2$ surface chemistry on GFET electrical characteristics. (a)** $R$-$V_G$ curves of the GFETs using hydrophobic (blue) and hydrophilic (red) $SiO_2$ thin films on Si(100) substrates. (It should be explained

in the Figure which of the red curves is forward and backward) No hysteresis occurs in graphene on hydrophobic $SiO_2$, hence the forward and backward sweep coincide. (**b**) Schematic diagram explaining the mechanism behind the hysteresis of the $R$-$V_G$ curve; the interface dipole layer plays a crucial role.

The next thing to be discussed is the comparison of Dirac voltages of the GFETs, where the resistivity ($R$) is the highest and the Fermi level is considered to coincide with the Dirac point, by relating the results obtained by 3D nano-ESCA, as described in the previous subsection. To relate with the 3D nano-ESCA measurements, we used the $R$-$V_G$ curve in the forward sweep for the GFET using a hydrophilic $SiO_2$ thin film. The reason for this choice is as follows. The $R$-$V_G$ curves were measured by sweeping $V_G$ from –30 V to +30 V (forward sweep) and later from +30 V to –30 V (backward sweep); it was stopped at –30 V, after which 3D nano-ESCA measurements at $V_G = 0$ V were carried out. Therefore, these measurements can be regarded to occur during a forward sweep. The Dirac voltage of the GFET using a hydrophilic $SiO_2$ thin film in the forward sweep is more positive than that recorded using a hydrophobic $SiO_2$ thin film. This indicates the graphene channel in the GFET using a hydrophilic $SiO_2$ thin film in the forward sweep is more hole-doped than that using a hydrophobic $SiO_2$ thin film. This result is consistent with the pinpoint C $1s$ core level spectra of the graphene channel, which indicate the binding energy shift toward lower energy on a hydrophobic $SiO_2$ thin film due to hole doping as shown in Fig. 2, although we must consider adsorbed molecules other than water, such as $O_2$, during the $R$-$V_G$ measurements[12,14]. These changes in the $R$-$V_G$ curves are thus explained by the modulation in the electronic states of graphene channels in terms of their interface chemistry with $SiO_2$ gate oxides, which was described in the previous subsection as demonstrated through 3D nano-ESCA.

### 3.4. Influence of $SiO_2$ Surface Chemistry near the Metal Contact

Surprisingly, $SiO_2$ surface chemistry also exerts an influence on the electronic states near the interface with the metal contact, which is also a key component in GFET. One of the consequences of such metal-contact influence is the formation of a charge transfer region (CTR)[13,40], which can extend up to a width of 1 μm in the GFET using hydrophilic $SiO_2$ as the gate oxide[22]. The CTR is supposed to be formed due to the disappearance of the density of states (DOS) near the Dirac point in graphene. Unfortunately, however, the influence of $SiO_2$ surface chemistry on the CTR is still unclear.

To clarify the influence of $SiO_2$ surface chemistry on the electronic states of the graphene channel near the interface between graphene and contact metal, we performed spatially resolved C $1s$ core level spectroscopy near the metal contact using 3D nano-ESCA. The spatial variation in the binding energy of graphene, which reflects the change in doping (work function), on hydrophilic and hydrophobic $SiO_2$ thin films is shown in Fig. 7. It can be inferred that across the entire measured range, the binding energy of graphene on a hydrophilic $SiO_2$ thin film is smaller than that on a hydrophobic $SiO_2$ thin film. This means that graphene on a hydrophilic $SiO_2$ thin film is more positively charged, compared to that on a hydrophobic $SiO_2$ thin film. The value of binding energy (~ 284.45 eV), which is very close to that of neutral graphene[27], indicates that the graphene channel is negligibly doped when a hydrophobic SiO₂ thin film was used, in agreement with the data in Figure 2.

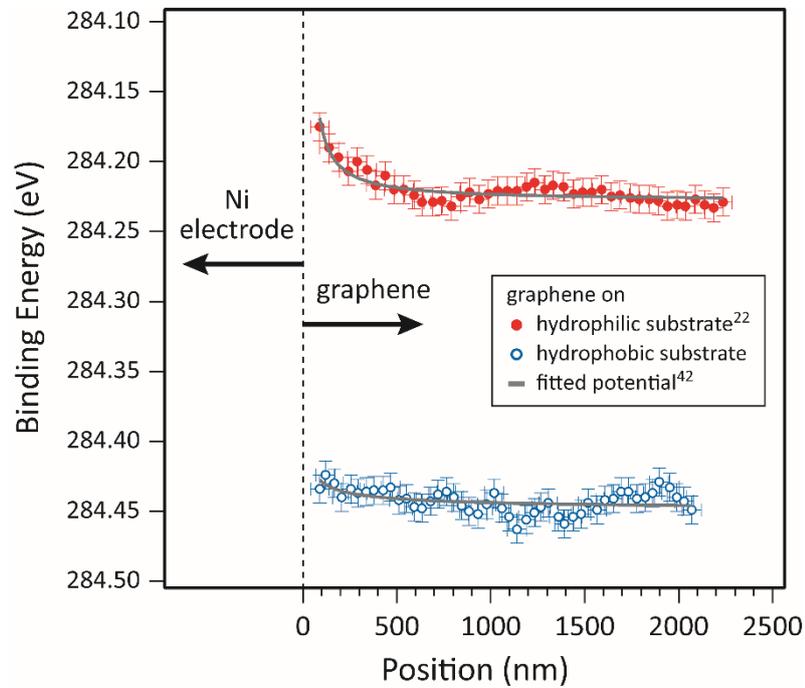

**Figure 7 │ Influence of SiO₂ surface chemistry on the potential variation across the graphene-contact metal interface.** Change in the C1*s* core level of graphene peak across the graphene/Ni interface in GFETs using hydrophilic (red) and hydrophobic (blue) SiO₂ thin films as the gate oxides. For comparison, data on hydrophilic SiO₂ thin films is sourced from ref. 22.

In sharp contrast to the hydrophobic SiO₂ thin film, the binding energy of graphene on hydrophilic SiO₂ becomes smaller near the contact metal as shown in Fig. 7. The binding energy shift originates from local charge density as shown in Fig. 2(c), so the results which display the spatial distribution of the binding energy shift in Fig. 7 can be interpreted as a direct measurement of the screening potential in graphene. Sub-micron CTR formation is detected in the graphene channel on the hydrophilic substrate, as reported in previous studies[22]. The determining factor in CTR formation is, in principle, supposed to be the charge transfer between materials of different work functions[13], 4.5 eV for graphene and 5.4 eV for Ni[41], which thermodynamically equilibrate the graphene/Ni system[22].

To explain the difference in the screening potential of the hydrophilic and hydrophilic substrates, we performed theoretical estimations of the screening potential according to the Thomas-Fermi approach proposed by Khomyakov *et al.*[42]. For a single layer of graphene, they described the screening potential in terms of the charge density in an ungated condition as follows

$$V(x) = \mu_F + \text{sign}(\sigma)\sqrt{\left|\frac{V_B|V_B|}{x/l_s} - \mu_F|\mu_F|\right|} \tag{3}$$

where $x$ is the distance from the metal/graphene contact edge, $l_s = \hbar v / \pi \alpha |V_B|$ is a scaling length, $V_B = V_{B1} + V_{B2}$, $\hbar v = 6.05 \text{ eV} \cdot \text{Å}$, and $\alpha = e^2 / 4\pi\varepsilon_0\kappa\hbar v = 2.38/\kappa$ is the fine-structure constant in graphene. $\kappa$ is the effective dielectric constant. $V_{B1}$ and $V_{B2}$ are boundary potential constants, which can be written as

$$V_{B1} = \frac{1}{4}(W - W_G), \qquad V_{B2} = \frac{\pi}{4}(W_M - W_G), \tag{4}$$

where $W_G$ is the work function of free-standing graphene (4.5 eV), $W_M$ is the work function of the contact metal layer (5.4 eV; Ni in this case), and $W$ is the work function of the graphene-covered metal. The parameter $\beta$ depends on the contact geometry and $\beta = \pi$ when a distance $x$ is large enough compared to $d$, a thickness of the contact metal ($x >> d$), where $d \sim 25$ nm in this case. $W$ was evaluated using density functional theory (DFT) calculations[43]. In the case of graphene on Ni(111), the conical dispersion in the graphene band is destroyed by strong graphene-metal bonding interactions[44]. However, in our process, the resistant residue prevents chemisorption between graphene and the Ni contact. Later, we can refer the value of an Au contact, which shows physisorption with graphene, and has a work function ($\sim 5.4$ eV) similar to that of a Ni contact. For large graphene-metal separations due to resistant residues, $W–W_G \sim 0.4$[43]. Subsequently, we obtained $\kappa \sim 1.8 \pm 0.9$ for graphene on a hydrophobic substrate and $\kappa \sim 77 \pm 4$ for graphene on a hydrophilic substrate[1] by curve fitting to the measured points in Fig. 7 using eq. (3) with $V_B \sim 0.325$. If we can neglect polarization effects at the graphene channel, the effective dielectric constant $\kappa$ is given by the average of the dielectric constant of $SiO_2$ ($\sim 3.9$ eV) and that of the vacuum due to the image effect[5], i.e., $\kappa \sim 2.5$. This value is close to the experimentally obtained value on the hydrophobic substrate. The large value of $\kappa$ on the hydrophilic substrate is due to the polarization of the water layer, which has a large dielectric constant at the graphene/substrate interface. Lacking spatial resolution, the interface dipole layer, of the order of several nanometers, cannot be detected in our system. However, the screening potential changes moderately at large values of $\kappa$ and we can detect spatial shifts in the screening potential by 3D nano-ESCA with a spatial resolution of ~100 nm. Therefore, the difference in the potential variation between the hydrophilic and hydrophobic substrates is caused by the difference in the effective dielectric constants, rather than the presence/absence of CTR. Although further theoretical investigation with quantum chemistry is required, we believe that the positive charging of graphene due to interactions with a hydrophilic $SiO_2$ thin film may assist graphene-Ni interactions, which in turn increases the amount of hole-doping in graphene near Ni, assuming that charge transfer occurs through bonding between graphene and Ni in the wide region which is larger than an interfacial dipole layer region.

---

[1] The estimated value of the effective dielectric constant, $\kappa$, is different from our previous study in ref. 22 because we adopted the undoped limit of Eq. (3) as a rough approximation. However, our argument that $\kappa$ shows larger value on a hydrophilic substrate than typical value of graphene's $\kappa$ ($\sim 2.5$) is consistent. The chemical potential $\mu_F$, in other words, the doping level of the graphene on the hydrophilic substrate is evaluated compared to the hydrophobic case in this study, so now we can use Eq. (3), which is more general fitting function than a previous study.

## 4. Conclusions

In summary, a combination of 3D nano-ESCA and device characteristics enabled us to quantitatively elucidate that $SiO_2$ surface chemistry as well as the metal contacts determine the electronic states of graphene channels and consequently, the GFET device performance. By using samples in a device geometry and layer arrangement, we observe a gate voltage induced reversal of the interface dipole orientation. The results obtained will serve as the basis for a quantitative understanding of the GFET operation mechanism, which will help in the realization of high-performance graphene-based devices.

## Acknowledgments

This work was supported by the Japan Society for the Promotion of Science (JSPS) through its "Funding Program for World-Leading Innovative R&D on Science and Technology (FIRST Program)" and "Grant-in-Aid for Scientific Research B (Grant Number: 15K17463)," the Ministry of Education, Culture, Sports, Science, and Technology (MEXT) through its "Grant-in-Aid for Scientific Research on Innovative Areas (Grant Number: 26107503)," Japan Science and Technology Agency (JST) through its "Core Research for Evolutional Science and Technology (CREST)," and "Materials Research by Information Integration" Initiative project and PRESTO (Grant Number: JPMJPR17NB), and the Research Program for CORE lab of "Dynamic Alliance for Open Innovation Bridging Human, Environment and Materials" in "Net-work Joint Research Center for Materials and Devices." This work was performed at the Synchrotron Radiation Research Organization, University of Tokyo (Proposal Nos. 7402 for 2009–2011, 2012-2014, 2015-2017, and 7435 for 2012).